
\catcode`\@=11
\font\tenmsx=msxm10
\font\sevenmsx=msxm7
\font\fivemsx=msxm5
\font\tenmsy=msym10
\font\sevenmsy=msym7
\font\fivemsy=msym5
\newfam\msxfam
\newfam\msyfam
\textfont\msxfam=\tenmsx  \scriptfont\msxfam=\sevenmsx
  \scriptscriptfont\msxfam=\fivemsx
\textfont\msyfam=\tenmsy  \scriptfont\msyfam=\sevenmsy
  \scriptscriptfont\msyfam=\fivemsy

\def\hexnumber@#1{\ifnum#1<10 \number#1\else
 \ifnum#1=10 A\else\ifnum#1=11 B\else\ifnum#1=12 C\else
 \ifnum#1=13 D\else\ifnum#1=14 E\else\ifnum#1=15 F\fi\fi\fi\fi\fi\fi\fi}

\def\msx@{\hexnumber@\msxfam}
\def\msy@{\hexnumber@\msyfam}
\mathchardef\square="0\msx@03
\catcode`\@=\active

\vsize=22.5cm
\hsize=14.5cm
\def\sk { \vskip .6cm }
\def\capit#1 {\vfill \eject \sk \centerline {\nbc #1}  \vskip 1cm }
\def\newpar#1 { \vskip 1.5cm \line {\nbb #1 \hfill}  \vskip .4cm}
\tolerance=1600

\overfullrule=0pt
$  $
\vskip1.5truein
\centerline {\bf The phase of scalar field driven wormholes }
\centerline {\bf at one loop in the path integral formulation }
\centerline {\bf for Euclidean quantum gravity}
\bigskip
\centerline{Alberto Carlini$^{(1)}$ and Maurizio Martellini$^{(2)}$}
\smallskip
\centerline{\it (1) International School for Advanced Studies, SISSA,}
\centerline{\it Strada Costiera 11, I-34014 Trieste, Italy}
\centerline{\it (2) Department of Physics, University of Milan,}
\centerline{\it Via Celoria 16, I-20133, Milan, Italy}
\centerline{\it and}
\centerline{\it I.N.F.N., University of Pavia, Pavia, Italy}
\vskip1truein
\centerline {\bf Abstract}
We here calculate the one-loop approximation to the Euclidean Quantum Gravity
coupled to a scalar field around the classical Carlini and Miji\'c
wormhole solutions.
The main result is that the Euclidean partition functional $Z_{EQG}$ in the
``little wormhole'' limit is real.
Extension of the CM solutions with the inclusion of a bare cosmological
constant to the case of a sphere $S^4$ can lead to the elimination of the
destabilizing effects of the scalar modes of gravity against those of the
matter.
In particular, in the asymptotic region of a large 4-sphere, we recover the
Coleman's $\exp \left (\exp \left ({1\over \lambda_{eff}}\right )\right )$ peak
at the effective cosmological constant $\lambda_{eff}=0$, with no phase
ambiguities in $Z_{EQG}$.
\vskip2truein
\vfil
\eject
{\bf 1. Introduction.}
\medskip
As it is well known, the Einstein-Hilbert action for the Euclidean Quantum
Gravity (QG) is not bounded from below.
This problem is particularly important for the physics of wormholes (see,
e.g., the Coleman's mechanism [1] for the suppression of the cosmological
constant $\Lambda$), since one has to compute a probability distribution of the
kind $\sim \exp \left (Z_{EQG}\right )$, where $Z_{EQG}$ is the Euclidean QG
partition function.
In this scheme, at the level of a tree expansion approximation for $Z_{EQG}$,
one can have various ``types'' of saddle points.
Even if the four sphere $S^4$ is usually taken as the dominant saddle point,
in a theory of interacting wormholes one can easily imagine that the wormhole
solutions also give a contribution to $Z_{EQG}$.
Then, it is important to determine whether $Z_{EQG}$ is a real number
or if it shows a phase ambiguity, as it has recently been debated by Polchinski
[2] and Mazur and Mottola [3].
The problem of the eventual existence of a phase in $Z_{EQG}$ could be
interesting by itself in the context of a theory for the Euclidean QG, since
a complex $Z_{EQG}$ would give rise to a free energy $F_{EQG}\sim -\ln \left
(Z_{EQG}\right )$ which is also complex, and this would be the signal of a
quantum instability (in other words, there would not be a stable ground
state for the quantum theory).

Recently, Carlini and Miji\'c (CM)[4] have shown that, beside the other known
solutions in the wormhole class, one can also construct an infinite number
of scalar field driven wormholes which represent the analytic continuation  of
closed expanding Robertson Walker (RW) universes.
Moreover, in [4] it has been shown that this new class of solutions exists
(at the classical level, $\hbar \rightarrow 0$) only provided  that the
gravitational and matter ``sectors'' undergo an ``orthogonal'' analytical
continuation to the Euclidean regime.
It is essential to stress that the Euclidean CM solutions are
different from those considered in [2] (which are $S^4$ spheres), since
the CM ansatz is represented by wormholes (with topology $R\times S^3$)
which are driven by the interaction of gravity with an additional explicit
matter contribution.
It is also interesting to note that the CM analytic continuation leading to a
nonstandard relative sign between the gravity and the matter Euclidean
actions exactly reproduces, in the specific case $\gamma_e=2$, the axionic
matter wormhole solutions previously found by Giddings and Strominger [5].

The main goal of this letter is to consider the one-loop expansion around
these CM solutions.
We find that, for ``little'' CM wormholes, $Z_{EQG}\in {\cal {R^+}}$,
thus removing the phase ambiguity described by [2].

It can also be shown that, extension of the CM solutions by the inclusion of a
bare cosmological constant, and assuming that the ground state is given by a
4-sphere $S^4$ with matter included, leads to the persistence of the Coleman
double exponential peak at the effective cosmological
constant equal to zero, with no phase ambiguities in $Z_{EQG}$.
\bigskip
{\bf 2. A review of the CM classical wormhole solutions.}
\medskip
CM considered asymptotically flat gravitational instantons (AFGI) which are
analytical continuations of closed expanding solutions in classical general
relativity (GR) and showed that wormholes are no more exotic than Friedmann
solutions [4].

First they studied the case with bulk matter sources (whose relevance is
a geometrical one) subject to the equation of state
$p=(\gamma -1)\rho$ ($p, \rho$ are the pressure and the energy density).
AFGIs exist only for matter sources that obey the strong energy condition
($\gamma > {2\over 3}$).
There are infinitely many instantons which are analytic continuation of
a classical closed universe solution in standard GR with maximum radius,
and explicit solutions can be written for $\gamma =const$ and homogeneous
and isotropic (RW) models.
The Euclidean metric is :
$$
ds_e^2 = a^{4-3\gamma} d\tau_e^2 + a^2 d\vec {x}^2 ~~,
{}~~~~~a(\tau_e)=\left [a_0^{3\gamma - 2} + \left ({{2-3\gamma} \over 2}
\right )^2 \tau_e^2 \right ]^{1/(3\gamma-2)} ~~.\eqno(2.1)
$$
with the lapse function $N_e^2=a^{4-3\gamma}$.
The wormhole neck is $a_0$ (at $\tau_e =0$, where it connects to the closed RW
universe at its maximum expansion), which is an arbitrary length scale
characterizing the solutions.
The wormhole metric is conformally equivalent
to an asymptotically flat geometry.
In particular, the cases $\gamma =2, \gamma ={4\over 3}$ respectively
correspond to the wormhole geometries of [5] and [6].

Another important result in [4] is the construction of the field theoretical
models that lead to these geometrical wormhole solutions.
In this context, it is shown that, for a fairly general case of a spatially
homogeneous minimally coupled scalar field $\phi$ with potential term $V$, this
is possible {\it only} if
one performs an asymmetric transition to the Euclidean regime : $N\rightarrow
\pm iN_e$ in the gravitational sector, but $N\rightarrow \mp iN_e$ in the
matter field sector, where $N, N_e$ are the lapse functions in the two regimes.
In the case of the Hamiltonian formalism, this is equivalent to using
the following effective Euclidean action :
$$
S_e = \pm 2\pi^2 \left [ \mu^2\int dt_e ~N_e
\left ( - {{a \dot {a}_e^2} \over {N_e^2}}
- a \right )  + \mu^2 t_e^2
- \int d\tau_e ~N_e a^3 \left ( {1\over 2}{\dot \phi^2 \over N_e^2}
 + V \right )  \right ] ~~.\eqno(2.2)
$$
The second term is due to regularization of the boundary term for gravity,
evaluated for flat Euclidean metric, $a=t_e$, and $\mu^2={3\over 8\pi G}$.
Now the equations of state in the two regimes are, in general, different, and
subject to the constraint $\gamma +\gamma_e=4$.
For ${2\over 3}<\gamma_e<{10\over 3}$ the wormhole solutions are the
continuation of RW closed expanding universes, while for $\gamma_e>{10\over 3}$
they create expanding inflationary universes  (see [4]).

The $\phi$ trajectory for the RW, $\gamma_e =const$ case has been determined
as :
$$
\phi (\tau_e)={2 \over {3\gamma_e -2}} \sqrt {\gamma_e} \mu
\arctan \left [ {{3\gamma_e -2} \over {2a_{0}^{(3\gamma_e -2)/2}}}
\tau_e \right ]
{}~~.\eqno(2.3)
$$
$$
\dot {\phi}_e = \mu \left ( \gamma_e a_{0}^{3\gamma_e -2} \right )^{1/2}a^{2-3
\gamma_e}   ~~.\eqno(2.4)
$$
The solutions have a nontrivial potential term :
$$
V(\phi_e)=\mu^2 {{\gamma_e -2} \over {2a_{0}^2}}
\left | \cos \left ( {{3\gamma_e -2} \over {2 \sqrt {\gamma_e} \mu}}
\phi_e \right ) \right |^{6\gamma_e /(3\gamma_e -2)} ~~.\eqno(2.5)
$$
In the asymptotic region ($a\rightarrow \infty$) the scalar field terms
all vanish ($\dot \phi \rightarrow 0, V\rightarrow 0$).
The manifest periodicity of $V$ and $\tau_e$ with respect to $\phi_e$ is
used to show that these spacetime wormholes have in fact a compact
topology $S^1\times S^3$.
Substitution of equations (2.1), (2.4) and (2.5) in (2.2) gives for the total
action the result :
$$
\eqalignno{
S_e =  &\pm {{3\pi} \over {2G}} a_{0}^2
{{\gamma_e - 2} \over {3\gamma_e -2}} \int_{0}^{+\infty} d\theta ~
\left [\cosh \theta \right ]^{2(4-3\gamma_e)/(3\gamma_e -2)} \cr
&\pm {{3\pi} \over G}a_0^2 {{2^{-4/(3\gamma_e -2)}} \over
{3\gamma_e -2}} \left [3(\gamma_e -2)
\exp \left ( 2{{4-3\gamma_e} \over {3\gamma_e -2}} \theta \right ) +
\pi \gamma_e \exp \left ( 3{{2-\gamma_e} \over {3\gamma_e -2}} \theta
\right ) \right ]_{|\infty} ~~. &(2.6)\cr}
$$
where a possible boundary term for the matter part of the action has been
included, and the metric has been parametrized as $a=a_0[\cosh \theta]
^{2/(3\gamma_e-2)}$.
For $\gamma_e =2$, this formula reproduces the [5] value, $S_e[2]=\pm {3\pi^2
a_o^2\over 4G}$.
The naive semiclassical amplitude for the wormhole geometries is finite for
$\gamma_e\geq 2$.
\bigskip
{\bf 3. Quantization of the classical solutions by the ``background'' field
method.}
\medskip
The partition function for EQG minimally coupled to matter ($Z_{EQG}$) is
usually defined as a functional integral :
$$
Z_{EQG}={\int [d ~g][d~ \Phi]\over V_{GC}}~e^{- S[g, \Phi]}\eqno(3.1a)
$$
$$
S=S_G+S_M\eqno(3.1b)
$$
where $S_G$ is the Euclidean gravitational part of the action (with bare
cosmological constant $\lambda$):
$$
S_G = - {1 \over {16\pi G}} \int_{\cal {M}} d^4x ~ \sqrt {g} (R-2\lambda) -
{1 \over {8\pi G}} \int_{\partial \cal {M}}d^3x~\sqrt {h} (K-K_o) ~~.\eqno(3.2)
$$
($K$ is the trace of the second fundamental form on the boundary of the
manifold
${\cal {M}}$, regularized by a similar term evaluated on flat spacetime,
$K_o$).
The coordinate group volume $V_{GC}$ compensates for the general coordinate
overcounting :
$$
V^{-1}_{GC}=\Delta_{FP}e^{-S_{GF}}\eqno(3.3)
$$
where $S_{GF}$ is a gauge fixing term and $\Delta_{FP}$ its consequent
Fadeev-Popov determinant (see, e.g. [2]).
$S_M$ represents the matter part of the action for a generic field $\Phi$.

The quantum theory at the one-loop order is studied in the ``background field
method'' by the expansion
of the quantum fields around some classical field configuration
(see, for instance, [7, 8]).
Namely, one expands the gravitational field and the matter field as:
$$
g_{\mu \nu}(x)=\hat g_{\mu \nu}(x)+2kh_{\mu \nu}(x)\eqno(3.4a)
$$
$$
\Phi (x)=\hat \phi (x)+\theta (x)\eqno(3.4b)
$$
($k^2={3\over \mu^2}$) where a ~$\hat {}$~  denotes a quantity evaluated on the
classical backgrounds ($\hat g_{\mu \nu}, \hat \phi$)
and $h_{\mu \nu}(x)$ and $\theta(x)$ are the quantum fluctuations (of order
$O(\hbar)$).
Therefore, the path integral in eq. (3.1) becomes :
$$
Z_{EQG}=N~\int [d ~h][d~ \theta]~\Delta_{FP}~e^{- S[\hat g +h, \hat \phi
 +\theta ]}\eqno(3.5)
$$
where $N$ is a normalization factor.

One then expands the action in a functional Taylor series about the classical
background (see also [8]):
$$
\eqalignno
{S[\hat g +h, \hat \phi +\theta]&\simeq S_o[\hat g, \hat \phi]+\int_{\cal
{M}}d^4 x~\left ({\delta S\over \delta g_{\mu \nu} (x)}\biggr\vert_{\hat g ,
\hat
\phi}h_{\mu \nu}(x) +{\delta S\over \delta \Phi (x)}\biggr\vert_{\hat g , \hat
\phi}
\theta (x)\right )+~\cr
&+~ {1\over 2!} \int_{\cal
{M}}d^4 xd^4 y~\biggl (h_{\mu \nu}(x){\delta^2 S\over \delta g_{\mu \nu} (x)
\delta g_{\rho \sigma} (y)}\biggr\vert_{\hat g , \hat \phi}
h_{\rho \sigma}(y)+\cr
&+2h_{\mu \nu}(x){\delta^2 S\over \delta g_{\mu \nu} (x)
\delta \Phi (y)}\biggr\vert_{\hat g , \hat \phi}\theta (y)
 +\theta (x){\delta^2 S\over \delta \Phi (x)\delta
\Phi (y)}\biggr\vert_{\hat g , \hat \phi}\theta (y)\biggr )~+..&(3.6)\cr}
$$
Now, the terms linear in the quantum fluctuations are zero because of the
equations of motion :
$$
{\delta S\over \delta \Phi (x)}\biggr \vert_{\hat g , \hat \phi}=
{\delta S\over \delta g_{\mu \nu} (x)}\biggr \vert_{\hat g , \hat \phi}=0
\eqno(3.7)
$$
The terms of the second order in the fluctuations (the Hessian of $S$) give
the quantum one-loop correction to the semiclassical theory (2.2).

Our basic problem here is to investigate the reality of $Z_{EQG}$.
Then, using standard `tricks' in the evaluation of the functional integral,
this amounts in studying only the phases coming from the Fredholm functional
determinants due to the ``full propagators'' for $h_{\mu \nu}$ and $\theta$.
Now, if one momentarily ignores the matter fluctuations, the critical fact
(observed by [2]) is that the determinant coming from the Gaussian integral
around the saddle point of the action will, in general, contain factors of
$i$ because the action $S$ of Euclidean gravity is unbounded from below.
Polchinski explicitly considered the Coleman's model [1] for the solution to
the cosmological constant problem.
Coleman's saddle point for $\lambda >0$ is a large 4 sphere of radius
$r=\sqrt {3\over \lambda k^2}$, and has action:
$$
S_o(\hat g)\simeq -{3\over 8G^2\lambda}\eqno(3.8)
$$
The sum over disconnected spheres, at semiclassical level, is given by a
probability distribution $\exp \left (Z_{EQG}\right )\simeq \exp \left (\exp
\left ( {3\over 8G^2\lambda}\right )\right )$, which appears to be infinitely
peaked at $\lambda =0$ (actually here $\lambda$ is the effective fully
renormalized
cosmological constant, which has become a dynamical variable due to the effects
of wormholes, see, e.g. [9]).

However, at quantum level, one has also to take into account the corrections
given by the field fluctuations, i.e. by the Hessian ${\cal {H}}$ of eq. (3.6):
$$
\int [d~h] e^{-h~{\cal {H}}~h}\simeq N~(Det ~{\cal {H}})^{-1/2}\eqno(3.9)
$$
In particular, if the Hessian has some negative eigenvalues, one should
rotate each corresponding eigenfunction in the complex plane by a factor of
$i$,
introducing a phase in $Z_{EQG}$ at one loop level.
Since ${\cal {H}}$ is almost completely negative definite around Coleman's
saddle point,
the prescription suggested by [2] is to globally rotate the Weyl parts of the
gravitational fluctuations $\phi \rightarrow i\phi$ (which gives a Jacobian
$J=1$) and then to rotate back
the eigenfunctions corresponding to the positive and zero eigenvalues of ${\cal
{H}}$ (in the number of $N_+$ and $N_0$).
Furthermore, Polchinski has shown that there is no phase ambiguity coming from
the Fadeev-Popov $\Delta _{FP}$.
This turns out as a global phase in front of $Z_{EQG}$ of the kind, in four
dimensions :
$$
i^{N_++N_0}=i^6=-1\eqno(3.10)
$$
which, therefore, would destroy Coleman's argument.
We incidentally note that the request of dropping out the negative (and zero)
eigenvalues of ${\cal {H}}$ also corresponds to the standard procedure for the
$\zeta$-
function regularization of $Det~{\cal {H}}$ (see [7]).
\bigskip
{\bf 4. The one-loop approximation to the CM wormhole solutions.}
\medskip
Let us now specialize the formalism introduced in the previous Section to the
case of the CM wormhole solutions.
We expand the total action , eq. (2.2) with the {\it plus} sign, around the
wormhole background fields $\hat g_{\mu \nu}(x),\hat \phi
(x)$ according to equations (3.4) and (3.6), and then we decompose the
gravitational
fluctuations into a symmetric, traceless tensorial part and a trace (Weyl)
scalar part (see [11]) as :
$$
h_{\mu \nu}=\phi_{\mu \nu} ~+~\hat g_{\mu \nu}\phi\eqno(4.1)
$$
where $\hat g^{\mu \nu}\phi_{\mu \nu}=0$ and $\hat g^{\mu \nu}h_{\mu \nu}=h=4
\phi $.

We first calculate the Hessian of the total action (3.1b) with respect
to the metric fluctuations $h_{\mu \nu}$.
The second order variation of the gravitational part of the action (given by
eq. (3.2), with $\lambda =0$) turns out as :
$$
\eqalignno {S_{G,2}={1\over 2}h_{\mu \nu}{\delta^2 S\over \delta g_{\mu \nu}
\delta g_{\rho \sigma} }\biggr\vert_{\hat g , \hat \phi}h_{\rho \sigma}&=\int
d^4x ~
\sqrt {\hat g}~ \biggl [{1\over 2}\tilde h_{\mu \nu}\left (-\hat g_{\mu \rho}
\hat g_{\nu \sigma}\hat \square +2\hat R_{\mu \rho}\hat g_{\nu \sigma}-
2\hat R_{\mu \rho \nu \sigma}\right )h^{\sigma \rho}\cr
&-\hat \nabla^{\rho}\tilde h_{\rho \mu}\hat \nabla^{\sigma}\tilde h_{\sigma}^
{\mu}-2\tilde h^{\mu \rho}\left (\hat R_{\rho \sigma}-{1\over 4}\hat g_{\rho
\sigma}\hat R \right )h^{\sigma}_{\mu} \biggr ]&(4.2)\cr}
$$
where we have defined (see [8]) :
$$
\tilde h_{\mu \nu}=h_{\mu \nu} -{1\over 2}\hat g_{\mu \nu}h\eqno(4.3)
$$
We therefore add a de Donder gauge fixing term [12] :
$$
S_{GF}=\int d^4x~\sqrt {\hat g}~\hat \nabla^{\rho}\tilde h_{\rho \mu}\hat
\nabla^{\sigma}\tilde h_{\sigma}^{\mu}\eqno(4.4)
$$
for which eq. (4.2) becomes :
$$
\eqalignno {
S_{G,2}+S_{GF}={1\over 2}\int d^4x~\sqrt {\hat g}~& (-\phi_{\rho \sigma}
\hat \square \phi^{\rho \sigma}+4\phi \hat
\square \phi -2\phi^{\mu \nu}\hat
R_{\mu \rho}\phi^{\rho}_{\nu}\cr
&-2\phi^{\mu \nu}\hat R_{\mu \rho \nu \sigma}
\phi^{\rho \sigma}+\phi^{\rho \mu}\hat R \phi_{\rho \mu})&(4.5)\cr}
$$
As for the Fadeev-Popov determinant $\Delta_{FP}$ (see eq. (3.5)) coming from
the gauge fixing, we follow [2] in accepting that it is the modulus
of this determinant which appears in $Z_{EQG}$, and therefore  $\Delta_{FP}$
does not affect the global phase of $Z_{EQG}$.

For the classical background of CM wormhole solutions, one can
express the curvature tensor $\hat R_{\mu \rho \nu \sigma}$ and the Ricci
scalar $\hat R$ in the following form :
$$
\hat R_{\mu \rho \nu \sigma}={\hat R\over 6}(\hat g_{\mu \sigma}
\hat g_{\rho \nu}-\hat g_{\mu \nu}\hat g_{\rho \sigma})+{1\over 2}
(\hat R_{\mu \nu}\hat g_{\rho \sigma}+\hat R_{\rho \sigma}\hat g_{\mu \nu}-
\hat R_{\mu \sigma}\hat g_{\rho \nu}-\hat R_{\rho \nu}\hat g_{\mu \sigma})
\eqno(4.6a)
$$
$$
\hat R=3(4-3\gamma )a_o^{3\gamma_e-2}a^{-3\gamma_e}\eqno(4.6b)
$$
Substitution of these explicit saddle point solutions into eq. (4.5) gives :
$$
S_{G,2}+S_{GF}=\int d^4x~\sqrt {\hat g}~\left (
\phi^{\mu \nu}\left (-{\hat \square \over 2}+(4-3\gamma_e)a_o^{3\gamma_e-2}
a^{-3\gamma_e}\right ) \phi_{\mu \nu}+2\phi \hat \square \phi \right
)\eqno(4.7)
$$
Let us now consider the Euclidean matter part of the total action, which will
be taken as :
$$
S_M=-\int_{\cal {M}}d^4x~\sqrt {g}\left ({\partial_{\mu}\Phi \partial^{\mu}\Phi
\over 2}+V\right )\eqno(4.8)
$$
where the scalar field $\Phi$ is held fixed on the boundary $\partial {\cal
{M}}$ and we choose a homogeneous classical solution $\hat \phi =\hat \phi
(\tau)$.
It is easy to show that the Gaussian fluctuations of $S_M$ with respect to
$\hat g_{\mu \nu}$ are given by :
$$
\eqalignno {S_{M,2}&={1\over 2}h_{\mu \nu}{\delta^2 S\over \delta g_{\mu \nu}
\delta g_{\rho \sigma} }\biggr\vert_{\hat g , \hat \phi}h_{\rho \sigma}=\cr
&=k^2\int d^4x ~
\sqrt {\hat g}\left [\phi^{\mu \nu}\left ({\dot {\hat \phi}^2
\over 2\hat N^2}+V(\hat \phi)\right )\phi_{\mu \nu}-4\phi V(\hat \phi)
\phi -2\phi^{0\sigma}\dot {\hat \phi}^2\phi^0_{\sigma}\right ]&(4.9)\cr}
$$
The same result can be obtained in the case
where the
momentum of the scalar field is held fixed on the boundary $\partial {\cal
{M}}$
[4].

Obviously, the only contribution to the Hessian of the total action $S$
with respect to the matter field fluctuations comes from $S_M$, and we can
write it as :
$$
{1\over 2}\theta {\delta^2 S\over \delta \Phi \delta
\Phi }\biggr\vert_{\hat g , \hat \phi}\theta ={1\over 2}\int d^4x~\sqrt {\hat
g}~\theta \left [\hat \square -{\delta^2 V\over \delta \Phi \delta
\Phi }\biggr\vert_{\hat g , \hat \phi}\right ]\theta\eqno(4.10)
$$
Then we can use the explicit expressions for the wormhole solutions given
in Section (2) for $\hat \phi, \dot {\hat \phi}$ and $\hat V(\hat \phi)$.
We first note that, according to eq. (2.5), the classical value $\hat V\dot =V(
\hat \phi)$
of the potential $V$ can be expanded as a power series in
$\hat \phi$.
If one assumes that the unknown potential $V(\Phi)$ can also be expanded as a
power series in $\Phi$, one is allowed to make the substitution :
$$
{\delta^2 V\over \delta \Phi \delta
\Phi }\biggr\vert_{\hat g , \hat \phi}={\delta^2 \hat V(\hat \phi)\over \delta
\hat \phi \delta \hat \phi }\biggr\vert_{\hat g , \hat \phi}\eqno(4.11)
$$
{}From eq. (2.5) we find :
$$
{\delta^2 \hat V(\hat \phi)\over \delta \hat
\phi \delta \hat \phi }\biggr\vert_{\hat g , \hat \phi}
={3\over 2\mu^2}V(\hat \phi)\left [(3\gamma_e+2)\tan^2
\left (
{3\gamma_e-2 \over 2\mu \sqrt {\gamma_e}}\hat \phi \right )-(3\gamma_e-2)\right
]\eqno(4.12)
$$
An explicit calculation of the phase coming from the determinants
of the
Hessian for the gravitational and matter field quantum fluctuations can be
made in two different regions of the CM classical wormhole solutions.

We will first consider the asymptotically flat region of the wormhole, i.e.
we will specialize to the limit :
$$
{\tau\over a_o^{(3\gamma_e-2)/2}}\gg 1\eqno(4.13)
$$
This bound can be achieved in particular for small $a_o$, which is the ``little
wormhole'' approximation.
In this region, we can reasonably approximate the classical values of ${\dot{
\hat \phi}^2\over \hat N^2}$ (eq. (2.4) and (2.1), $\hat V$ (eq. (2.5)),
${\delta^2 \hat V\over \delta \hat \phi^2}$
(eq. (4.12)), the curvature tensor and the second term in eq. (4.7) by zero.
Therefore, in this limit, the second order variation of $S_M$ with respect
to $g_{\mu \nu}$ (eq. (4.9)) vanishes, and we are left with the following final
formula for the one-loop contribution to $Z_{EQG}$ :
$$
\eqalignno {
S_{TOT,2}&\simeq \int d^4x~\sqrt {\hat g}~\left (
\phi^{\mu \nu}\left (-{\hat \square \over 2}\right )
 \phi_{\mu \nu}+2\phi \hat \square \phi +{1\over 2}
\theta\hat \square \theta
\right )\cr
&\dot =\int (\phi^{\mu \nu}A\phi_{\mu \nu}+\phi B\phi +\theta C\theta)
{}~\sqrt {\hat g}~d^4x &(4.14)\cr}
$$
Let us better clarify this last result.
It comes out from the discussion of Section 2 that one has the freedom to
choose
the global sign in front of the Euclidean  action (3.1b).
Our choice, which corresponds to define an Euclidean negative definite matter
action, suggests that, for the CM wormhole, matter behaves like the conformal
degrees of freedom in the metric field.
But this is actually the consequence of two facts: first, the necessity of
the nonstandard choice in the relative sign of the matter and gravitational
sectors of the Euclidean action for the existence of the CM wormholes and,
second, the adoption of the usual normalization prescription
for the complex rotation of the conformal factor of gravity.
Despite this unusual definition, the (negative) Euclidean matter
action evaluated at the classical (instantonic) level is well behaved, in
particular it is bounded from below, at least in the parameter range
$\gamma_e>{4\over 3}$.

For our purpose, the next step is to determine the spectrum of the Hermitian
differential operators $A, B, C$ acting
on $\phi_{\mu \nu}, \phi$ and $\theta$, since we have formally :
$$
\int [d ~fluct]e^{-S_{TOT,2}}\sim \left (Det ~A~\cdot Det ~B~\cdot Det~C
\right )^{-{1\over 2}}\eqno(4.15)
$$
Of course, the above functional determinant must be regularized in some way.

To determine the eigenvalues of the background Laplacian operator in a four
dimensional Riemannian manifold, we remember that the action of this operator
on a symmetric second rank tensor $T_{\mu \nu}$, in the above limit of
vanishing curvature, is the same as (see [11])
$$
\hat \square T_{\mu \nu}\simeq -\hat \square_L T_{\mu \nu}\eqno(
4.16)
$$
where $\square_L$ is the generalized Laplacian introduced by Lichnerowicz [13].
It is well known (see, for instance, [14]) that the spectrum of $\square_L$ on
a compact manifold is positive semi-definite.
Since the background of CM wormhole solutions may be compactified (see
Section (2)), and this spectral property of $\square_L$ also holds for the
contraction of $T_{\mu \nu}$ into a scalar, we conclude that :
$$
\eqalignno
{spectrum~(A) \geq 0&\cr
spectrum~ (B) \leq 0&&(4.17)\cr
spectrum ~(C) \leq 0&\cr}
$$
Therefore, in order to avoid problems connected with the evaluation of
$Z_{EQG}$ and for the consistency of the saddle point method here used,
it is necessary
to drop out the eigenfunctions of the operators corresponding to the zero and
negative eigenvalues.
Let us first take care of the negative eigenspace.
The operator $A$ has no troubles with the negative eigenvalues.
To deal with the $B$ and $C$ operators, on the contrary, one
follows [2] in rotating the entire
integration over both $\phi$ and $\theta$ in the Euclidean path integral (EPI)
 by a factor of $i$
(whose corresponding Jacobians can be neglected).
Then one rotates back
the eigenfunctions corresponding to zero eigenvalues (which are of finite
number $N_{0}(B)=N_0(C)$ and depend on the choice of the topology of the
spacetime).
If we strictly followed this prescription, i.e. if we rotated $\phi \rightarrow
\pm i \phi$ and $\theta \rightarrow \pm i\theta$ (with the {\it same} relative
sign), we would obtain for the global phase in $Z_{EQG}$
the following result :
$$
i^{(N_0(B)+N_0(C))}=(-1)^{N_0(B)}=-1\eqno(4.18)
$$
where for a closed manifold (as the Euclidean continued CM wormholes)
$N_0(B)=N_0(C)=1$, corresponding to the constant eigenfunctions [15].
This would give the same distressing result as in [2].

Therefore, our fundamental attitude is to define the prescription for
the Wick rotation of the scalar terms in the one-loop expansion of $Z_{EQG}$
as :
$$
\phi \rightarrow + i\phi \eqno(4.19a)
$$
$$
\theta \rightarrow - i\theta \eqno(4.19b)
$$
which, therefore, gives a global real phase :
$$
(+ i)^{N_0(B)}(- i)^{N_0(C)}=+1\eqno(4.20)
$$
After dropping out the zero modes following the standard way [16], the
Euclidean quantum gravity partition function obtained in this way is real
with no phase ambiguities.

Let us note that the assumption of a negative
definite Euclidean matter action is not in contradiction with the
claims made in [2].
The key point is that the ansatz considered by [2] is that of a $S^4$ sphere,
which is radically {\it different} from the CM ansatz, where one has a
different topology ($R\times S^3$), the scalar field driven wormholes, and
a different Euclidean action.

It is also interesting to note that, in this new context, the classical
prescription of [4] might receive a more serious motivation and justification.
In a certain sense, we could say that our prescription can be seen as
the extension of that proposed by [10] for the conformal degrees of
freedom of the gravitational metric, where we have just added a rule for the
case where also a matter contribution is present in the action.
One can expect that it is just the quantum theory that makes this difference
between gravity and matter.

To conclude this Section, we briefly mention about another possible
 approximation around the CM classical background wormhole solutions.
The approximation is to work in the background spacetime region near the
wormhole ``neck'', i.e. at :
$$
{\tau\over a_o^{(3\gamma_e-2)/2}}\ll 1\eqno(4.21)
$$
Again, the above bound for fixed $\tau$ can be achieved for large $a_o$, i.e.
in
the ``giant wormhole'' limit.
In this case, using eq. (4.7), (4.9), and (4.10), it is almost straightforward
to see that eq. (4.14) is substituted by :
$$
S_{TOT,2}=\int d^4x~\sqrt {\hat g}~ (\phi^{i \nu}A^{\prime}\phi_{i \nu}+
\phi B^{\prime}\phi+\theta C^{\prime}\theta+\phi^{0\sigma}D^{\prime}\phi^0
_{\sigma})\eqno(4.22)
$$
where we have set :
$$
\eqalignno
{A^{\prime}=&-{\hat \square \over 2}+{1\over a_o^2}&(4.23a)\cr
B^{\prime}=&2\left (\hat \square -3{(\gamma_e-2)\over a_o^2}\right )&(4.23b)\cr
C^{\prime}=&{1\over 2}\left (\hat \square +{3(3\gamma_e-2)(\gamma_e-2)\over
4a_o^2}\right )&(4.23c)\cr
D^{\prime}=&-{\hat \square \over 2}+{(1-6\gamma_e)\over a_o^2}&(4.23d)\cr}
$$
Unfortunately, the spectrum of these operators is not as easy to be found as in
the previous case.
This is not surprising, since in this large $a_o$ approximation the dynamical
role of matter couplings affects the small scale structure (i.e. the one-loop
approximation) by a sort of a ``classical'' gravitational dressing induced by
these ``giant wormholes''.
We hope to return on this point in a future work.
\bigskip
{\bf 5. Generalization.}
\medskip
Now, it would be interesting to extend the discussion to the case corresponding
to the Coleman ansatz for the suppression of the cosmological constant
$\lambda$.
This would imply assuming that the ground state of the Euclidean theory is
given
by a 4-sphere $S^4$ including the matter contributions.
For the Euclidean CM solutions this corresponds taking $\gamma_e=0$.

Recently, the Coleman's theory has been questioned by Unruh and Hawking [17],
which pointed out that the peak at $\lambda =0$ might be the obvious
consequence
of the unboundedness of the Euclidean gravitational action , and by Polchinski
[2], who claimed the existence of a complex phase in front of $Z_{EQG}$.
We will demonstrate that, also in this case, adoption of the CM definition
of the Euclidean path integral can lead to the
``annihilation'' of the destabilizing effects of the gravitational field modes
against the matter ones.
Here we will not give all the details, which will be instead discussed in a
future publication [18].

The idea is to extend the CM solutions to the case where a bare cosmological
constant $\lambda$ is included in the gravitational action, as we can read from
eq. (3.2).
In the approximation $a_o^2\lambda \ll 1$, the CM classical Euclidean
geometries
can be generalized to have a scale factor [18] :
$$
a(\lambda)=a(0)\cdot (1+\lambda a_0^2f(x))\eqno(5.1)
$$
where $a(0)$ is the CM solution (eq. (2.1)) with $\lambda =0$, and we have put
$x\dot ={\vert 3\gamma_e-2\vert \tau \over 2a_0^{(3\gamma_e-2)/2}}$.
These solutions, in the case $\gamma_e>{2\over 3}$, still represent a wormhole
geometry, which now connects a (Lorentzian) Tolman universe at its maximum
radius to a (Lorentzian) de Sitter universe at its minimum size.
Another interesting case is when one considers $\gamma_e=0$, which essentially
corresponds to the topology of a 4-sphere $S^4$ with {\it matter contribution
included}, and for which it is found that :
$$
f(x)=-{1\over 6}\cdot {1\over (1+x^2)}\eqno(5.2)
$$
If we now insist in taking the CM-definition of the Euclidean action
also in the case $\gamma_e=0$, the explicit form of the
classical potential $\hat V(\lambda)$ (which is now independent of $\hat \phi$,
since, from eq. (2.4), $\hat \phi =const$, for $\gamma_e=0$) can be written as
:
$$
\hat V(\lambda)\simeq -{\mu^2\over a_0^2}+O(\lambda^2)\eqno(5.3)
$$
and we can define an effective cosmological constant as :
$$
\lambda_{eff}\dot =\lambda -{3\over \mu^2}\hat V(\lambda)\eqno(5.4)
$$
{}From eq. (5.1) and (5.4), it is then possible to explicitly compute the
classical value of the action (3.1b), which now becomes :
$$
\hat S_e=2\int d^4x~\sqrt {\hat g(\lambda)}\left [-{\mu^2\over a_0^2}(1+x^2)
(1-2\lambda a_0^2f(x))+{\mu^2\over 3}\lambda_{eff}\right ]\eqno(5.5)
$$
Repeating a similar kind of calculations as those made in Section 4, it is easy
to show that the (generalized) Gaussian fluctuations of the total action
(3.1b),
in the one-loop approximation and $\gamma_e=0$ case, come out as :
$$
\eqalignno {
S_{TOT,2}&= \int d^4x~\sqrt {\hat g(\lambda)}~\left [
\phi^{\mu \nu}\left (-{\hat \square (\lambda)\over 2}+{4\over a_0^2}-\lambda_
{eff}\right )
 \phi_{\mu \nu}+2\phi \left (\hat \square (\lambda)+2\lambda_{eff}\right )\phi
+{1\over 2}\theta\hat \square (\lambda)\theta
\right ]\cr
&\dot =\int (\phi^{\mu \nu}A^{\prime \prime}\phi_{\mu \nu}+\phi B^{\prime
\prime
}\phi +\theta C^{\prime \prime}\theta)
{}~\sqrt {\hat g(\lambda)}~d^4x &(5.6)\cr}
$$
and, for the scaling behaviour, we have :
$$
\hat \square (\lambda)\biggr \vert_{\gamma_e=0}\sim {1\over a(\lambda)^2}\hat
\nabla^2_3\sim {(1+x^2)\over a_0^2}\hat \nabla^2_3\eqno(5.7)
$$
where $\hat \nabla_3^2$ is the 3-D Laplacian for $S^3$.
Now, again we can assume to work in the asymptotic limit $x\rightarrow \infty$,
which is the case, for instance, if $\tau$ is large.
In this limit, using eq. (5.7), we can easily check that the spectrum of the
differential operators $A^{\prime \prime}$ and $C^{\prime \prime}$ remains the
{\it same} as that of the operators $A$ and $C$ (eq. (4.14)), while $B^{\prime
\prime}$ can have at most one positive (or zero) eigenvalue, if we work in the
range :
$$
0\leq \lambda_{eff}\leq {1\over a_0^2}(1+x^2)\eqno(5.8)
$$
Therefore, if one just assumes $\lambda_{eff}\in {\cal {R^+}}$ (which is the
same as in [1]), and follows the same discussion as in Section 4, it can be
seen that the $S^4$ generalization of the CM classical solutions gives, at the
one-loop approximation, $Z_{EQG}\in {\cal {R^+}}$.
Moreover, for $x\rightarrow \infty$, now we have $f(x)\rightarrow -1/6x^2$ and
$\lambda_{eff}\rightarrow 3/a_0^2$.
{}From eq. (5.5) we then obtain the classical action :
$$
\hat S_e\simeq -{\mu^2\over \lambda_{eff}}\eqno(5.9)
$$
As a consequence, the Coleman double exponential peak at
$\lambda_{eff}=0$ survives in the ansatz of the $\lambda$-enlarged CM solutions
for a large $S^4$ sphere with the inclusion of a non-trivial matter content,
without the extra phase in $Z_{EQG}$, found by Polchinski [2].
\bigskip
{\bf 6. Conclusions.}
\medskip
As we have shown, the one-loop Euclidean QG path integral for the CM ansatz
is free of the phase ambiguities which affect the usual standard $S^4$-wormhole
theories.
What we have computed is just the one-loop Euclidean correction to the CM
classical solutions which represent a ``particular'' kind of matter interacting
with gravity, but which possess the global geometric features of the wormholes.

In other words, we are not questioning here the necessity of extending the CM
definition of the Euclidean action to every matter coupled to the Einstein
gravity.
Rather, we are simply pointing out that, in the absence of a set of
``reconstruction axioms'' for interacting QG, one is actually left free to
define different Euclidean theories.
Obviously, these possible different theories will also imply, in principle,
different dynamics.
\bigskip
{\bf Acknowledgements.}
\medskip
We would like to thank Prof. G. Veneziano, Prof. E. Massa, Prof. Monti Bragadin
and Dr. M. Miji\'c for useful discussions.
This work has been supported by the Italian Ministero per l'Universita' e la
Ricerca Scientifica e Tecnologica.
\vfil
\eject

\def\tindent#1{\indent\llap{#1}\ignorespaces}
\def\ref{\par\hang\tindent}


{\bf References:}
\medskip
\ref{$1.$ }S. Coleman, {\it Nucl. Phys.} {\bf B310}, 643, (1988).
\ref{$2.$ }J. Polchinski, {\it Phys Lett.} {\bf B219}, 251, (1989).
\ref{$3.$ }P.O. Mazur and E. Mottola, {\it Nucl. Phys.} {\bf B341}, 187,
(1989).
\ref{$4.$ }A. Carlini and M. Miji\'c, {\it SISSA} preprint {\it 91A}, (1990).
\ref{$5.$ }S. Giddings and A. Strominger, {\it Nucl. Phys.} {\bf B306}, 890
(1988).
\ref{$6.$ }S.W. Hawking, {\it Phys. Rev.} {\bf D37}, 904, (1988).
\ref{$7.$ }S.W. Hawking, {\it Comm. Math. Phys.} {\bf 55}, 133, (1977).
\ref{$8.$ }S.M. Christensen and M.J. Duff, {\it Nucl. Phys.} {\bf B170}, 480,
(1980).
\ref{$9.$ }I. Klebanov, {\it PUPT} preprint {\it 1153}, (1989); S.W. Hawking,
{\it DAMTP} preprint {\it R-89-13}, (1989).
\ref{$10.$ }G.W. Gibbons, S.W. Hawking and M.J. Perry, {\it Nucl. Phys.}
{\bf B138}, 141, (1978).
\ref{$11.$ }G.W. Gibbons and M.J. Perry, {\it Nucl. Phys.} {\bf B146}, 90,
(1978).
\ref{$12.$ }G. 't Hooft and M. Veltman, {\it Ann. Inst. H. Poincare'} {\bf A20}
, 69, (1975).
\ref{$13.$ }G. De Rham, {\it ``Variet\'es Differentiables''}, (Hermann \& C.
ed., Paris, 1955).
\ref{$14.$ }I. Chavel, {\it ``Eigenvalues in Riemannian geometry''}, (Academic
Press, Orlando, 1984).
\ref{$15.$ }S.W. Hawking, {\it ``General Relativity, an Einstein Centenary
Survey''}, (ed. Hawking and Israel, Cambridge Univ. Press, 1979).
\ref{$16.$ }P. Ramond, {\it ``Field theory, a modern primer''}, (ed. The
Benjamin/Cummings Publ. Comp., 1981).
\ref{$17.$ }W.G. Unruh, {\it Phys. Rev.} {\bf D40}, 1053, (1989); S.W. Hawking,
{\it Mod. Phys. Lett.} {\bf A5}, 453, (1990).
\ref{$18.$ }A. Carlini, {\it SISSA} preprint, in preparation.
\vskip1.5truein
\vfill
\eject
\bye